\documentclass{article}

\usepackage{amssymb}
\usepackage{amsmath}

\begin{document}

\title{Quantum-classical correspondence via a deformed kinetic operator}

\author{
R. A. Mosna\thanks{Instituto de Matem\'atica, Estat\'\i stica e
Computa\c{c}\~ao Cient\'\i fica, Universidade Estadual de
Campinas, CP 6065, 13081-970, Campinas, SP, Brazil. E-mail
address: mosna@ime.unicamp.br}, I. P. Hamilton\thanks{Department
of Chemistry, Wilfrid Laurier University, Waterloo, Canada N2L
3C5. E-mail address: ihamilto@wlu.ca} \ and L. Delle
Site\thanks{Max-Planck-Institute for Polymer Research,
Ackermannweg 10, D 55021 Mainz Germany. E-mail address:
dellsite@mpip-mainz.mpg.de} }

\maketitle

\begin{abstract}

We propose an approach to the quantum-classical correspondence
based on a deformation of the momentum and kinetic operators of
quantum mechanics. Making use of the factorization method, we
construct classical versions of the momentum and kinetic operators
which, in addition to the standard quantum expressions, contain
terms that are functionals of the N-particle density. We show that
this implementation of the quantum-classical correspondence is
related to Witten's deformation of the exterior derivative and
Laplacian, introduced in the context of supersymmetric quantum
mechanics. The corresponding deformed action is also shown to be
related to the Fisher information. Finally, we briefly consider
the possible relevance of our approach to the construction of
kinetic-energy density functionals.

\medskip


\end{abstract}

\maketitle
\section{Introduction}
\label{sec intro}

Since the origins of quantum mechanics there has been interest in
the correspondence between quantum and classical mechanics which
has continued to the present \cite{qcc}. In his 1926 paper
\cite{Schr} Schr\"{o}dinger begins with the classical
Hamilton-Jacobi equation and then writes down a wavefunction
equation (now known as the Schr\"{o}dinger equation) without
making an explicit connection between the two. A general
connection was made by Van Vleck in his 1928 paper \cite{VanV} and
extended by Schiller \cite{Schi} who modifies the classical
Hamilton-Jacobi equation to obtain a quantum-like formulation of
classical mechanics. On the other hand, in his 1928 paper
\cite{Mad} Madelung begins with the wavefunction in polar form and
then writes down hydrodynamic equations to obtain a classical-like
formulation of quantum mechanics. This approach was extended by
Bohm \cite{Bohm} who explicitly introduces the quantum potential,
$Q$. One can think of the quantum-classical correspondence as
``switching off'' the quantum potential term in the modified
(quantum) Hamilton-Jacobi equation \cite{Holland} and this
approach was explicitly explored in Ref.~\cite{Ghose}. The $Q\to
0$ and (more usual) $\hbar\to 0$ approaches to the
quantum-classical correspondence are discussed in
Ref.~\cite{Holland} (see also Ref.~\cite{Bolpaper}).

In this paper we approach the quantum-classical correspondence at
the level of the equations of motion of an N-particle system.
Recall that, by expressing the wavefunction in polar form, the
Schr\"{o}dinger equation can be transformed into two
equations~\cite{Mad,Gosh}: a modified Hamilton-Jacobi equation in
which the quantum potential, $Q$, appears in addition to the
external potential, and a continuity equation. In this context,
one formally obtains the quantum equations of motion by
``switching on'' the quantum potential term in the classical
Hamilton-Jacobi equation. As we review in
section~\ref{sec.clasquan}, one may similarly think of
formally obtaining the classical equations of motion by
``switching on'' the quantum potential term in the Schr\"odinger
equation \cite{Holland}. Either way, $Q$ is tacitly regarded as an
additional {\em potential} term in these approaches.

In this work we follow an alternate approach, in which $Q$ is
incorporated in the {\em kinetic} term. This different perspective
results in a deformed kinetic operator which naturally motivates
the search for a deformed momentum operator. In section~\ref{sec
pcl}, making use of the factorization method \cite{factorref}, we
construct classical versions of the momentum and kinetic operators
and show that, in addition to the standard quantum expressions,
these operators contain terms which are functionals of the
N-particle density. We show that our classical version of the
momentum operator is equivalent to a certain classical momentum
component, introduced by Hall in \cite{Hall}, which gives the best
classical estimate that is compatible with simultaneous knowledge
of the position of the system. In section~\ref{sec pgen} we show
that our approach is formally related to Witten's deformation of
the exterior derivative and Laplacian \cite{Witten} (see also
\cite{Rapoport} and Appendix A of Ref.~\cite{Urs}). In
section~\ref{sec.fisher} we show that the deformed momentum
operator reproduces, at the action level, the quantization
procedure developed by Reginatto \cite{reg}, where the
Schr\"{o}dinger equation is derived from an
information-theoretical approach based on the principle of minimum
Fisher information \cite{fisher}. Our conclusions are presented in
section~\ref{sec conclusion}, where we also discuss how our work
is related to that of Hall and Reginatto~\cite{HR}, who approach
the quantum-classical correspondence by introducing certain
momentum fluctuations obeying an exact Heisenberg-type equality.
Finally, the possible relevance of our approach to the
construction of kinetic-energy density functionals (see
Refs.~\cite{jpa,prb} and references therein) is briefly considered
in the Appendix.

\section{Quantum-classical correspondence and the quantum potential}

\label{sec.clasquan}

In the hydrodynamic formulation of quantum mechanics the quantum
potential appears as an additional potential term in a modified
Hamilton-Jacobi equation, which defines a classical-like
description of quantum mechanics. In a complementary way, the
quantum potential also appears as an additional term in a modified
Schr\"{o}dinger equation which defines a quantum-like description
of classical mechanics \cite{Holland,Ghose}. In this section, we
briefly review these procedures.

\subsection{Classical-like description of quantum mechanics}
\label{sec.QM}

Consider a \emph{quantum} N-particle system described by a
wavefunction $\psi=\psi(\mathbf{r}_1,...,\mathbf{r}_N,t)$
satisfying the Schr\"{o}dinger equation
\begin{equation}
i\hbar\frac{\partial\psi}{\partial t}=-\sum_{k=1}^N
\frac{\hbar^{2}}{2m_k}\nabla_{\!k}^{2}\psi+V\psi,
\label{Schroedinger equation}
\end{equation}
where $m_k$ is the mass of the $k$th particle and $V$ includes
interparticle and external potentials. Writing
$\psi=\sqrt{\rho}e^{\frac{i}{\hbar}S}$, and expressing the
Schr\"{o}dinger equation in terms of $\rho$ and $S$
yields~\cite{Mad,Holland}

\begin{subequations}
\label{QHJ}
\begin{align}
\frac{\partial S}{\partial
t}+\sum_{k=1}^N\frac{(\boldsymbol{\nabla}_{\!k} S)^{2}}{2m_k}+V+Q
& =0,
\label{QHJ.a}\\
\frac{\partial\rho}{\partial t}+\sum_{k=1}^N
\boldsymbol{\nabla}_{\!k}\cdot\left( \rho
\frac{\boldsymbol{\nabla}_{\!k} S}{m_k}\right) & =0,
\label{QHJ.b}
\end{align}
\end{subequations}
where $Q$ is the \emph{quantum potential}
\[
Q=-\sum_{k=1}^N\frac{\hbar^{2}}{2m_k}\frac{\nabla_{\!k}^{2}\sqrt{\rho}}{\sqrt{\rho}}.
\]

The system of coupled equations (\ref{QHJ}) comprises a modified
Hamilton-Jacobi equation --- in which $Q$ appears in addition to
the classical potential $V$ --- and a continuity equation. In this
way, it is the quantum potential that introduces, in the context
of Eq.~(\ref{QHJ.a}), all non-classical effects of quantum
mechanics, such as superposition, interference and entanglement
\cite{Ghose,Gosh}.

\subsection{Quantum-like description of classical mechanics}

\label{sec.CM}

In this case, one starts from a \emph{classical} N-particle system
whose action function $S$ is governed by the usual Hamilton-Jacobi
equation $\frac{\partial S}{\partial t}+
\sum_{k=1}^N\frac{(\boldsymbol{\nabla}_{\!k} S)^{2}}{2m_k}+V=0$
\cite{Goldstein}. The associated (local) momentum is given by
$\mathbf{p}_k=m_k\frac{d\mathbf{r}_k}{dt}=\boldsymbol{\nabla}_{\!k}
S$. Consider a distribution function $\rho$ defining an ensemble
of such trajectories, and satisfying the continuity equation
$\frac{\partial\rho}{\partial t}+ \sum_{k=1}^N
\boldsymbol{\nabla}_{\!k}\cdot\left(
\rho\frac{\boldsymbol{\nabla}_{\!k} S}{m_k}\right) =0$. One can
then introduce the so-called ``classical wavefunction''
\cite{Holland,Ghose}
\begin{equation}
\psi_{cl}=\sqrt{\rho}e^{iS/\hbar}, \label{classical wavefunction}
\end{equation}
which can be shown to satisfy the modified Schr\"{o}dinger
equation
\begin{equation}
i\hbar\frac{\partial\psi_{cl}}{\partial t}= \left(
-\sum_{k=1}^N\frac{\hbar^{2}}{2m_k}\nabla_{\!k}^{2}+V\right)
\psi_{cl}-Q\psi_{cl}. \label{classical Schroedinger eq}
\end{equation}

Note that, because of the last term,
Eq.~(\ref{classical Schroedinger eq}) is \emph{nonlinear}.
In this way, the quantum potential term in
Eq.~(\ref{classical Schroedinger eq}) completely eliminates the
quantum characteristics of the usual (linear) Schr\"{o}dinger equation,
giving rise to a purely classical behavior, as described by the
Hamilton-Jacobi equation.

\section{Classical versions of momentum and kinetic operators}
\label{sec pcl}

The interpretation of $Q$ as an additional {\em potential} term in
the particle's equation of motion has far-reaching consequences,
as the hydrodynamical formulation of quantum mechanics shows.
Notwithstanding, both Eqs.~(\ref{QHJ.a}) and (\ref{classical
Schroedinger eq}) allow an alternative interpretation of $Q$ as a
deformation of the {\em kinetic} term in the corresponding
equation of motion. In particular, one can readily interpret the
additional term in Eq.~(\ref{classical Schroedinger eq}) as a
deformation of the kinetic operator in quantum mechanics,
\[
-\sum_{k=1}^N\frac{\hbar^{2}}{2m_k}\nabla_{\!k}^{2}\rightarrow
-\sum_{k=1}^N\frac{\hbar^{2}}{2m_k}\nabla_{\!k}^{2} - Q.
\]

This motivates the definition of a classical version of the
kinetic operator as
\begin{equation}
K_{cl}=-\sum_{k=1}^N\frac{\hbar^{2}}{2m_k}\nabla_{\!k}^{2} - Q.
\label{Kcl}
\end{equation}

This change of perspective motivates the search for a classical
version of the momentum operator to be associated with $K_{cl}$.
We now develop this idea.

In the spirit of the factorization method in quantum mechanics
\cite{factorref}, we assume a factorization of Eq.~(\ref{classical
Schroedinger eq}) as
\begin{equation}
\left(
-\sum_{k=1}^N\frac{\hbar^{2}}{2m_k}\nabla_{\!k}^{2}-Q\right) \psi=
\sum_{k=1}^N\frac{1}{2m_k}(-i\hbar\boldsymbol{\nabla}_{\!k}
-\mathbf{g}_k)\cdot
(-i\hbar\boldsymbol{\nabla}_{\!k}+\mathbf{g}_k)\psi,
\label{factor}
\end{equation}
where $\mathbf{g}_k$ is a function to be determined. Expanding
this expression yields
\[
\sum_{k=1}^N \frac{1}{m_k}\left(
i\hbar\boldsymbol{\nabla}_{\!k}\cdot\mathbf{g}_k+\mathbf{g}_k^{2}+
\hbar^{2}\frac{\nabla_{\!k}^{2}\sqrt{\rho}}{\sqrt{\rho}}
\right)\psi=0.
\]
This suggests that we choose a purely imaginary $\mathbf{g}_k$, of
the form $\mathbf{g}_k=i\boldsymbol{\alpha}_k$ (with
$\boldsymbol{\alpha}_k$ real) which yields
\[
\sum_{k=1}^N \frac{1}{m_k}\left(
\hbar\boldsymbol{\nabla}_{\!k}\cdot\boldsymbol{\alpha}_k+\boldsymbol{\alpha}_k^{2}-
\hbar^{2}\frac{\nabla_{\!k}^{2}\sqrt{\rho}}{\sqrt{\rho}}
\right)\psi=0.
\]
Now we note that
$\frac{1}{\sqrt{\rho}}\nabla_{\!k}^{2}\sqrt{\rho}=
\boldsymbol{\nabla}_{\!k}\cdot\left(
\frac{1}{\sqrt{\rho}}\boldsymbol{\nabla}_{\!k}\sqrt{\rho}\right)
-\boldsymbol{\nabla}_{\!k}\left( \frac{1}{\sqrt{\rho}}\right)
\cdot \boldsymbol{\nabla}_{\!k}\left( \sqrt{\rho}\right) =
\boldsymbol{\nabla}_{\!k}\cdot\left(
\frac{1}{\sqrt{\rho}}\boldsymbol{\nabla}_{\!k}\sqrt{\rho}\right) +
\left(
\frac{1}{\sqrt{\rho}}\boldsymbol{\nabla}_{\!k}\sqrt{\rho}\right)^{2}$.
Substituting in the above equation yields the following condition
on $\boldsymbol{\alpha}_k$:
\[
\sum_{k=1}^N \frac{1}{m_k}\left[
\hbar\boldsymbol{\nabla}_{\!k}\cdot\boldsymbol{\alpha}_k+\boldsymbol{\alpha}_k^{2}-
\hbar\boldsymbol{\nabla}_{\!k}\cdot\left(
\frac{\hbar}{\sqrt{\rho}}\boldsymbol{\nabla}_{\!k}\sqrt{\rho}\right)-
\left(
\frac{\hbar}{\sqrt{\rho}}\boldsymbol{\nabla}_{\!k}\sqrt{\rho}\right)^{2}
\right]\psi=0.
\]
This is immediately fulfilled by the choice
\[
\boldsymbol{\alpha}_k=\hbar\frac{\boldsymbol{\nabla}_{\!k}\sqrt{\rho}}{\sqrt{\rho}}=
\frac{\hbar}{2}\frac{\boldsymbol{\nabla}_{\!k}\rho}{\rho},
\]
with $k=1,\ldots,N$.

Bearing in mind Eq.~(\ref{factor}), the discussion above motivates
the definition of a classical version of the momentum operator as
\begin{equation}
\mathbf{P}_{\!cl}
=\mathbf{P}+i\frac{\hbar}{2}\frac{\boldsymbol{\nabla}\rho}{\rho},
\label{Pcl0}
\end{equation}
and correspondingly,
$\mathbf{P}_{\!cl}^{\dag}=\mathbf{P}-i\frac{\hbar}{2}\frac{\boldsymbol{\nabla}\rho}{\rho}$.
Here $\mathbf{P}=-i\hbar\boldsymbol{\nabla}$ is the usual
N-particle quantum mechanical momentum operator, with
$\mathbf{P}=(\mathbf{P}_1,\ldots,\mathbf{P}_N)$ and
$\boldsymbol{\nabla}=(\boldsymbol{\nabla}_1,\ldots,\boldsymbol{\nabla}_N)$.

It is crucial to note that the N-particle density $\rho$ in the
expressions of $\mathbf{P}_{\!cl}$ and $\mathbf{P}_{\!cl}^{\dag}$
is that associated with the wavefunction $\psi$ of the system (so
that $\rho=\psi^{\ast}\psi$), regardless of the function on which
$\mathbf{P}_{\!cl}$ and $\mathbf{P}_{\!cl}^{\dag}$ operate; the
notation $\mathbf{P}_{\!cl}^{\dag}$ is to be understood in this
context (see also below). Therefore, the operator
$\mathbf{P}_{\!cl}$ is a functional of the N-particle density and,
in particular, we can write (for the wavefunction of the system),
\[
\mathbf{P}_{\!cl}\psi=-i\hbar\left(\boldsymbol{\nabla}-
\frac{1}{2}\frac{\boldsymbol{\nabla}(\psi^{\ast}\psi)}{\psi^{\ast}\psi}\right)\psi,
\]
so that the action of $\mathbf{P}_{\!cl}$ is nonlinear in $\psi$.
In this way, the notation $\mathbf{P}_{\!cl}^{\dag}$ is not to be
interpreted as the Hermitian conjugate of a (standard) linear
operator in $\psi$. Nonetheless, $\mathbf{P}_{\!cl}$ and
$\mathbf{P}_{\!cl}^{\dag}$ are complementary in that
$\mathbf{P}_{\!cl}$ arises naturally from the exterior derivative
while $\mathbf{P}_{\!cl}^{\dag}$ arises naturally from its
coderivative (see section~\ref{sec pgen}). Also, the last term in
Eq.~(\ref{Pcl0}) can be interpreted as representing momentum
fluctuations (see section~\ref{sec conclusion}) so that, in a
certain sense, $\mathbf{P}_{\!cl}$ and $\mathbf{P}_{\!cl}^{\dag}$
are statistically conjugate.

The classical version of the kinetic operator of Eq.~(\ref{Kcl})
can then be expressed as
\begin{equation}
K_{cl}=\sum_{k=1}^N\frac{1}{2m_k}(\mathbf{P}_{\!cl}^{\dag})_k\cdot(\mathbf{P}_{\!
cl})_k. \label{Kcl2}
\end{equation}
In the limiting case of a one-particle system ($N=1$), this simplifies to
\[
K_{cl}=\frac{1}{2m}\mathbf{P}_{\!cl}^{\dag}\cdot\mathbf{P}_{\!
cl}.
\]

Note that $\mathbf{P}_{\!cl}$ plays the role of a classical
version of the momentum operator, in the sense that it is
associated with the kinetic term $K_{cl}$ appearing in the
modified Schr\"{o}dinger equation (\ref{classical Schroedinger
eq}). As Eq.~(\ref{classical Schroedinger eq}) is nonlinear, we see
that, as a matter of fact, it is mandatory for $\mathbf{P}_{\!cl}$
to be nonlinear in $\psi$.

It is interesting to note that the action of $\mathbf{P}_{\!cl}$
on the wavefunction $\psi$ of the system is given by
\begin{equation}
\mathbf{P}_{\!cl}\psi=\hbar\operatorname{Im}\left(
\frac{\boldsymbol{\nabla}\psi}{\psi}\right)\psi,
\label{cmo}
\end{equation}
as a straightforward calculation shows. Therefore, writing
$\psi=\sqrt{\rho}e^{iS/\hbar}$, we find
\begin{equation}
\mathbf{P}_{\!cl}\psi=\boldsymbol{\nabla}S \: \psi. \label{Pcl1}
\end{equation}
As discussed in section \ref{sec.CM}, given the ``classical
wavefunction'' $\psi_{cl}=\sqrt{\rho}e^{iS/\hbar}$, the quantity
$\boldsymbol{\nabla}S$ can be interpreted as the local momentum
associated with an ensemble of trajectories. This reinforces our
interpretation of $\mathbf{P}_{\!cl}$ as a classical version of
the momentum operator. Further justification on this point
can be given by noting that
\begin{equation}
\langle \mathbf{P} \rangle_{\psi} = \langle \mathbf{P}_{\!cl} \rangle_{\psi},
\label{average.momentum}
\end{equation}
where $\langle A \rangle_{\psi} = \int \psi^{\ast} A \psi~d\mathbf{r}_1...d\mathbf{r}_N$
is the expectation value of the operator $A$ (here in position representation)
in the state $\psi$.

From a different perspective, an equivalent definition of a
classical momentum, $\mathcal{P}_{\!cl}$, associated with a given wavefunction
$\psi$, was introduced in \cite{Hall}.
In these works, Hall defines a state-dependent decomposition of the momentum
observable $P$ into ``classical'' and ``nonclassical'' components
$P = \mathcal{P}_{\!cl} + \mathcal{P}_{\!nc}$. The classical component
$\mathcal{P}_{\!cl}(\mathbf{r})$ corresponds to the best possible estimate
of the momentum which is compatible with simultaneous knowledge of the
position of the system, and is given by \cite{Hall}
\begin{equation}
\mathcal{P}_{\!cl}(\mathbf{r})=\hbar\operatorname{Im}\left(
\frac{\boldsymbol{\nabla}\psi(\mathbf{r})}{\psi(\mathbf{r})}\right).
\label{cmf}
\end{equation}
In this way, the classical version of the momentum operator
$\mathbf{P}_{\!cl}$ of Eq.~(\ref{Pcl0}) is essentially equivalent
to the classical momentum function
$\mathcal{P}_{\!cl}(\mathbf{r})$ of Eq.~(\ref{cmf}). We also note
that Eq.~(\ref{average.momentum}) was already obtained in
\cite{Hall} with $\mathcal{P}_{\!cl}$ in the place of
$\mathbf{P}_{\!cl}$.

Let us now compare the action of $\mathbf{P}$ and
$\mathbf{P}_{\!cl}$ on some particular one-dimensional examples.
For a plane wave $\psi=e^{\frac{i}{\hbar}p_{0}x}$ ($p_{0}$
constant), we see that the actions of $\mathbf{P}_{\!cl}$ and
$\mathbf{P}$ coincide, so that
$\mathbf{P}_{\!cl}\psi=\mathbf{P}\psi=p_{0}\psi$. On the other
hand, it follows from Eq.~(\ref{Pcl0}) that any $\psi$ with a
nontrivial probability distribution $\rho=\psi^{\ast}\psi$ will
lead to different results under the action of $\mathbf{P}$ and
$\mathbf{P}_{\!cl}$. Consider for example a Gaussian wave packet
with width $\sigma$,
$\psi(x)=e^{-(x-x_{0})^{2}/(2\sigma)^{2}}e^{\frac{i}{\hbar}p_{0}
x}$, which corresponds to a state with uncertainties in position
and momentum minimizing the Heisenberg uncertainty principle. Note
that, although $\psi$ is as close we can get to a quantum state
with definite position $x_{0}$ and momentum $p_{0}$, it is not an
eigenstate of $\mathbf{P}=-i\hbar\partial_{x}$ and therefore does
not correspond to a state of definite momentum. On the other hand,
we readily see that $\mathbf{P}_{\!cl}\psi=p_{0}\psi$. This is a
consequence of the fact that $\mathbf{P}_{\!cl}$ only cares about
the local momentum (in the sense above) associated with $\psi$.

\section{Witten's approach to supersymmetric quantum mechanics}
\label{sec pgen}

Some time ago, Witten constructed a connection between
supersymmetric quantum mechanics and Morse theory \cite{Witten}
which has been very influential. Central to Witten's approach
\cite{Witten} is the deformation of the exterior derivative, $d$,
\begin{equation}
d\rightarrow d_{\lambda}=e^{-\lambda f}de^{\lambda f},\quad
\lambda\in\mathbb{R} \label{deform.deriv}
\end{equation}
and correspondingly, the deformation of its coderivative,
$\delta=d^{\dag}$, $\delta \rightarrow
\delta_{\lambda}=e^{\lambda f}\delta e^{-\lambda f}$ (see also
\cite{Rapoport} and Appendix A of Ref.~\cite{Urs}). We note that,
by choosing
$f=-\frac{1}{2}\ln\rho$, the action of $d_{\lambda}$ on scalar
functions (i.e., 0-forms) is given by
\begin{equation}
-i\hbar\boldsymbol{\nabla}^{(\lambda)}=-i\hbar\boldsymbol{\nabla}+
i\hbar\frac{\lambda}{2}\frac{\boldsymbol{\nabla}\rho}{\rho},
\end{equation}
where, as before, $\boldsymbol{\nabla}$ denotes the
$3N$-dimensional gradient and we employ a vector notation for the
outcome of $d$ and $d_{\lambda}$ when applied to a scalar
function.\footnote{This amounts to identifying $1$-forms with
vector fields via the isomorphism induced by the Euclidean metric
in $\mathbb{R}^{3N}$, so that the actions of $d$ and $d_{\lambda}$
on a scalar function $\varphi$ are mapped into
$\boldsymbol{\nabla}\varphi$ and
$\boldsymbol{\nabla}^{(\lambda)}\varphi$, respectively.}

This suggests (cf Eq.~(\ref{Pcl0})) that we define a deformed
momentum operator, $\mathbf{P}_{\lambda}$, by
\begin{equation}
\mathbf{P}_{\lambda}
=\mathbf{P}+i\lambda\frac{\hbar}{2}\frac{\boldsymbol{\nabla}\rho}{\rho},
\label{Pgen}
\end{equation}
and correspondingly, $\mathbf{P}_{\lambda}^{\dag}
=\mathbf{P}-i\lambda\frac{\hbar}{2}\frac{\boldsymbol{\nabla}\rho}{\rho}$.
Given the relationship between the factorization method
and supersymmetric quantum mechanics \cite{factorref},
this connection between $\mathbf{P}_{\lambda}$ and $d_{\lambda}$
is, in fact, not unexpected. Once again, it is crucial to bear
in mind that the density $\rho$ in Eq.~(\ref{Pgen}) is that for
the wavefunction $\psi$ of the system, regardless of the function
on which $\mathbf{P}_{\lambda}$ (and $\mathbf{P}_{\lambda}^{\dag}$)
operate and the notation $\mathbf{P}_{\lambda}^{\dag}$ is to be
understood in this context. Also, the same observations made for
$\mathbf{P}_{\!cl}$ concerning its nonlinearity in $\psi$ also
apply to $\mathbf{P}_{\lambda}$.
For $\lambda=0$, $\mathbf{P}_{\lambda}$ recovers the usual
quantum momentum while for $\lambda=1$, $\mathbf{P}_{\lambda}$
recovers the classical version, $\mathbf{P}_{\!cl}$, of the
previous section. As $\lambda$ increases from 0 to 1, one can
envisage a scenario in which quantum mechanics gradually assumes
classical effects.

Also central to Witten's approach \cite{Witten} is the deformation
of the Laplacian, $L$,
\begin{equation}
L\rightarrow L_{\lambda}=(d_{\lambda}+\delta_{\lambda})^{2}.
\label{deform.lap}
\end{equation}
$L_{\lambda}$ is the natural Laplacian corresponding to
$d_{\lambda}$ and $\delta_{\lambda}$. When restricted to scalar
functions, it is not difficult to show that $L_{\lambda}$
satisfies $\frac{\hbar^2}{2m}L_{\lambda}\bigr\rvert
\begin{minipage}[l]{2.3em}
\renewcommand{\baselinestretch}{0.5}
\tiny \vspace{2ex}
\textit{scalar}\\
\textit{functions}
\end{minipage}
=\frac{1}{2m}\mathbf{P}_{\lambda}^{\dag}\cdot\mathbf{P}_{\lambda}$.
This suggests (cf Eq.~(\ref{Kcl2})) that we define a deformed
kinetic operator, $K_{\lambda}$, by
\begin{equation}
K_{\lambda}=\sum_{k=1}^N\frac{1}{2m_k}(\mathbf{P}_{\lambda}^{\dag})_k\cdot(\mathbf{P}_{\lambda})_k.
\label{Kgen}
\end{equation}
In the limiting case of a one-particle system ($N=1$), this
simplifies to
\[
K_{\lambda}=\frac{1}{2m}\mathbf{P}_{\lambda}^{\dag}\cdot\mathbf{P}_{\lambda}.
\]
For $\lambda=0$, $K_{\lambda}$ recovers the usual quantum kinetic
operator, while for $\lambda=1$, $K_{\lambda}$ recovers its
classical version, $K_{cl}$, of the previous section.

We note that a related connection between the quantum potential and supersymmetry
was considered in \cite{Rapoport}, where the quantum potential is obtained from a
Riemann-Cartan-Weyl geometry and deformed Laplacians are associated with generators
of a family of diffusion processes.

\section{Deformed action and Fisher information}

\label{sec.fisher}

The Fisher information, $I_F$, was introduced in statistical analysis
as a measure of the intrinsic accuracy of an estimate \cite{fisher}.
Given an N-particle system with associated probability density $\rho$,
it has been shown \cite{reg} that $I_F$ is directly proportional to the
average of $Q$. For simplicity, we restrict ourselves to the case $N=1$,
where $I_F$ is given by
\[
I_F=\int\frac{1}{\rho}\left( \boldsymbol{\nabla}\rho\right)
^{2}d\mathbf{r}dt.
\]
We note that, in this case, the Fisher information and the
Weizs\"{a}cker term, $W$, in Density Functional Theory
\cite{YangParr} are directly proportional with $W = \frac{\hbar^2}{8m}I_F$.

In order to relate the deformed momentum operator $\mathbf{P}_{\lambda}$
to Fisher information, we recall that, from a field theoretical viewpoint,
the Schr\"{o}dinger equation can be derived from the action \cite{Holland}
\begin{equation}
\mathcal{S}= \int\left[ i\frac{\hbar}{2}\left(
\psi^{\ast}\frac{\partial\psi}{\partial t}-
\frac{\partial\psi^{\ast}}{\partial t}\psi\right) -
\frac{1}{2m}(\mathbf{P}\psi)^{\ast}\cdot(\mathbf{P}\psi)-V\psi^{\ast}\psi\right]
d\mathbf{r}dt, \label{Ssch}
\end{equation}
where $\mathbf{P}=-i\hbar\boldsymbol{\nabla}$ (as above).

When we substitute $\mathbf{P}\rightarrow\mathbf{P}_{cl}$ in
Eq.~(\ref{Ssch}) we obtain the classical action
\begin{equation}
\mathcal{S}_{cl}= \int\left[ i\frac{\hbar}{2}\left(
\psi^{\ast}\frac{\partial\psi}{\partial t}-
\frac{\partial\psi^{\ast}}{\partial t}\psi\right)-
\frac{1}{2m}(\mathbf{P}_{cl}\psi)^{\ast}\cdot(\mathbf{P}_{cl}\psi)-
V\psi^{\ast}\psi\right] d\mathbf{r}dt, \label{Scl1}
\end{equation}
and for $\psi=\sqrt{\rho}e^{\frac{i}{\hbar}S}$, this expression
becomes
\begin{equation}
\mathcal{S}_{cl}=-\int\rho\left( \frac{\partial S}{\partial t}+
\frac{\left( \boldsymbol{\nabla}S\right) ^{2}}{2m}+V\right)
d\mathbf{r}dt, \label{Scl2}
\end{equation}
which gives rise (upon variation of $\rho$ and $S$) to the
{\em classical} Hamilton-Jacobi equation $\frac{\partial S}{\partial
t}+\frac{(\boldsymbol{\nabla}S)^{2}}{2m}+V=0$, along with the
continuity equation $\frac{\partial\rho}{\partial t}+
\boldsymbol{\nabla}\cdot\left(\rho\frac{\boldsymbol{\nabla}S}{m}\right)
=0$.

It is then natural to consider the deformed action
\begin{equation}
\mathcal{S}_{\lambda}= \int\left[ i\frac{\hbar}{2}\left(
\psi^{\ast}\frac{\partial\psi}{\partial t}-
\frac{\partial\psi^{\ast}}{\partial t}\psi\right) -
\frac{1}{2m}(\mathbf{P}_{\lambda}\psi)^{\ast}\cdot(\mathbf{P}_{\lambda}\psi)-
V\psi^{\ast}\psi\right] d\mathbf{r}dt, \label{Slambda1}
\end{equation}
associated with the deformed momentum $\mathbf{P}_{\lambda}$
introduced above. For $\psi=\sqrt{\rho}e^{\frac{i}{\hbar}S},$ a
straightforward calculation yields
\begin{equation}
\mathcal{S}_{\lambda}= -\int\left[  \rho\left(  \frac{\partial
S}{\partial t}+ \frac{\left(  \boldsymbol{\nabla}S\right)
^{2}}{2m}+V\right)  + \xi~\frac{1}{\rho}\left(
\boldsymbol{\nabla}\rho\right)  ^{2}\right] d\mathbf{r}dt,
\label{Slambda2}
\end{equation}
with
\[
\xi=\frac{\hbar^{2}(1-\lambda)^{2}}{8m}.
\]
It is interesting to note that this equation is the starting point
of Ref.~\cite{reg} to show that the Schr\"{o}dinger equation may be
derived from an information-theoretical approach based on the
principle of minimum Fisher information.\footnote{Note that the
parameter $\xi$ in Eq.~(\ref{Slambda2}) is a function of $\lambda$
and is thus far free. To avoid confusion, we note that in
\cite{reg} the author uses a slightly different notation,
with our $\xi$ corresponding to his $\frac{\lambda}{m}$.}
This can be seen as follows. In Eq.~(\ref{Slambda2}), the Lagrange
multiplier $\xi$ imposes a constraint to the purely classical Hamilton-Jacobi
action (\ref{Scl2}) which enforces the minimization of $I_F$ for a
given $S$ \cite{reg2}. As a result, $\mathcal{S}_{\lambda}$ yields
--- upon variation of $\rho$ and $S$ --- the continuity equation
$\frac{\partial\rho}{\partial t}+ \boldsymbol{\nabla}\cdot\left(
\rho\frac{\boldsymbol{\nabla}S}{m}\right) =0$ (for any value of
$\xi$), along with the modified Hamilton-Jacobi equation
$\frac{\partial S}{\partial
t}+\frac{(\boldsymbol{\nabla}S)^{2}}{2m}+V+ \xi\left(
-4\frac{\boldsymbol{\nabla}^{2}\sqrt{\rho}}{\sqrt{\rho}}\right)
=0$ which, as we have seen, is equivalent to the Schr\"{o}dinger
equation for $\xi=\frac{\hbar^{2}}{8m}$ (in our case, this
corresponds to $\lambda=0$).

\section{Discussion and Conclusions}
\label{sec conclusion}

We have proposed an approach to the quantum-classical
correspondence via deformed momentum and kinetic operators. We
started by reviewing the role played by the quantum potential,
$Q$, in providing the quantum-classical correspondence at the
level of the equations of motion of an N-particle system. However,
instead of regarding $Q$ as an additional {\em potential} term, we
have regarded $Q$ as resulting from a deformation of the {\em
kinetic} term in the corresponding equation of motion. We
introduced a deformed momentum operator, $\mathbf{P}_{\lambda}$,
which is related to Witten's deformation of the exterior
derivative, and a deformed kinetic operator, $K_{\lambda}$, which
is related to Witten's deformation of the Laplacian. The deformed
momentum and kinetic operators each contain the standard quantum
mechanical expression and an additional term which is a functional
of the N-particle density.

We have shown that our approach leads to connections to the
factorization method, to Witten's approach to supersymmetric
quantum mechanics and to the Fisher information, and each of these
might lead to more general formalisms.

We note that from a different perspective, Hall and Reginatto
\cite{HR} approach the quantum-classical correspondence by
introducing momentum fluctuations which scale inversely with the
uncertainty in position, so that the assumption of an exact
uncertainty principle leads from the classical equations of motion
to the Schr\"{o}dinger equation. In this context, it may be seen
that the momentum fluctuations introduced in \cite{HR} are
captured, in our formalism, by the last term of Eq.~(\ref{Pcl0}).
It is interesting to note that, although Hall and Reginatto
approach the quantum-classical correspondence from a different
perspective, there are significant similarities to our approach in
that we also obtain a modified kinetic term and a classical
version of the quantum mechanical momentum, as discussed in
section~\ref{sec pcl}.

\bigskip
\noindent\textbf{Acknowledgments} We thank A. O. Bolivar, D. L.
Rapoport and Urs Schreiber for helpful comments. RAM acknowledges
FAPESP for financial support and IPH acknowledges funding from
NSERC.

\begin{appendix}
\section{A brief note on kinetic-energy density functionals}

As noted in section~\ref{sec.fisher}, our approach is related
to the Fisher information and this concept has given rise
to several applications in Density Functional Theory~\cite{Nagy,spd}.
Density Functional Theory is, in principle, an exact physical theory
(within the framework of the Born-Oppenheimer approximation); the
total energy and other properties are expressed as functionals of
the one-electron density which is the basic variable. We conclude by
suggesting a way in which our approach might be relevant to the
construction of kinetic-energy density functionals (see Refs.~\cite{jpa,prb}
and references therein).

We write the N-electron wavefunction as
$\psi=\sqrt{\rho}e^{\frac{i}{\hbar}S}$, where
$\rho=\rho(\mathbf{r}_1,...,\mathbf{r}_N)$ is the N-electron
density. Recall that $\mathbf{P}_{\!cl}$ in Eq.~(\ref{Pcl0})
contains the standard quantum mechanical expression and an
additional term which is a functional of the N-electron density.
For N particles of mass $m$ (which we can take as the electron
mass) our construction of the classical version of the kinetic
operator in Eq.~(\ref{Kcl2}) yields
$K_{cl}=\frac{1}{2m}\mathbf{P}_{\!cl}^{\dag}\cdot\mathbf{P}_{\!cl}$,
where the scalar product implicitly contains a sum over N
particles. On the other hand, a simple (and physically reasonable)
construction of a kinetic operator, also built from
$\mathbf{P}_{\!cl}$ and $\mathbf{P}_{\!cl}^{\dag}$, is
$K^\prime=\frac{1}{2m}\frac{\mathbf{P}_{\!cl}^{2}+
\mathbf{P}_{\!cl}^{\dag 2}}{2}$ (which also implicitly contains
sums over N particles). A straightforward calculation yields
\begin{equation}
K^\prime = K- \frac{\hbar^{2}}{8m}
\frac{[\boldsymbol{\nabla}\rho(\mathbf{r}_1,...,\mathbf{r}_N)]^2}{[\rho(\mathbf{r}_1,...,\mathbf{r}_N)]^2}.
\label{expk}
\end{equation}
This then yields
\begin{equation}
\langle K \rangle_{\psi}=\langle K^\prime \rangle_{\psi}+
\frac{\hbar^{2}}{8m}\int
\frac{(\boldsymbol{\nabla}\rho(\mathbf{r}_1,...,\mathbf{r}_N))^2}{\rho(\mathbf{r}_1,...,\mathbf{r}_N)}
d\mathbf{r}_1...d\mathbf{r}_N, \label{weisz}
\end{equation}
where $\langle A \rangle_{\psi}$ represents (as before) the expectation
value of a given operator $A$ in the state $\psi$.

The average kinetic energy is thereby expressed as the sum of two
terms: the average of a kinetic-energy term constructed from the
classical version of the momentum operator
plus a 3N-dimensional analog of the Weizs\"{a}cker term. The
kinetic-energy density functional is typically expressed as a
linear combination of Thomas-Fermi and Weizs\"{a}cker terms
\cite{YangParr,prb} and it would be interesting if, upon reduction
to an expression in which the one-electron density is the basic
variable, Eq.~(\ref{weisz}) were to yield an expression of similar
form. We note that Kohout has recently considered a related
reduction for the 3N-dimensional quantum potential \cite{kohout}.
\end{appendix}

\end{document}